\pdfoutput=1

\documentclass[11pt,twoside,a4paper,cmspaper,final,collab]{cms-tdr}
\def\svnVersion{281855}\def\cmsCernNo{2013-037}\def\cmsCernDate{\today}\def\cmsMessage{Published in Physics Letters B as \href{http://dx.doi.org/10.1016/j.physletb.2015.03.017}{\doi{10.1016/j.physletb.2015.03.017.}}}

\begin{document}\cmsNoteHeader{SUS-14-009}

\hyphenation{had-ron-i-za-tion}
\hyphenation{cal-or-i-me-ter}
\hyphenation{de-vices}
\RCS$Revision: 277938 $
\RCS$HeadURL: svn+ssh://svn.cern.ch/reps/tdr2/papers/SUS-14-009/trunk/SUS-14-009.tex $
\RCS$Id: SUS-14-009.tex 277938 2015-02-18 14:11:53Z alverson $
\newlength\cmsFigWidth
\ifthenelse{\boolean{cms@external}}{\setlength\cmsFigWidth{0.85\columnwidth}}{\setlength\cmsFigWidth{0.4\textwidth}}
\ifthenelse{\boolean{cms@external}}{\providecommand{\cmsLeft}{top}}{\providecommand{\cmsLeft}{left}}
\ifthenelse{\boolean{cms@external}}{\providecommand{\cmsRight}{bottom}}{\providecommand{\cmsRight}{right}}

\newcommand{\htt}{\ensuremath{H_{\mathrm{T}}}\xspace}
\newcommand{\st}{\ensuremath{S_{\mathrm{T}}}\xspace}
\newcommand{\nj}{\ensuremath{N_{\text{jets}}}\xspace}
\newcommand{\nbj}{\ensuremath{N_{\text{b-jets}}}\xspace}
\newcommand{\sgo}{\ensuremath{\widetilde{\mathrm S}}\xspace}
\newcommand{\ggo}{\ensuremath{\widetilde{\chi}_{1}}\xspace}
\cmsNoteHeader{SUS-14-009}
\title{Search for stealth supersymmetry in events with jets, either photons or leptons, and low
  missing transverse momentum in pp collisions at 8\TeV}

\date{\today}

\abstract{
  The results of a search for new physics in final states with jets, either photons or leptons, and
  low missing transverse momentum are reported. The study is based on a sample of proton-proton
  collisions collected at a center-of-mass energy $\sqrt{s} = 8\TeV$ with the CMS detector in 2012.
  The integrated luminosity of the sample is 19.7\fbinv. Many models of new physics predict the
  production of events with jets, electroweak gauge bosons, and little or no missing transverse
  momentum. Examples include stealth models of supersymmetry (SUSY), which predict a hidden sector
  at the electroweak energy scale in which SUSY is approximately conserved. The data are used to
  search for stealth SUSY signatures in final states with either two photons or an oppositely
  charged electron and muon. No excess is observed with respect to the standard model expectation,
  and the results are used to set limits on squark pair production in the stealth SUSY framework.
}

\hypersetup{%
pdfauthor={CMS Collaboration},%
pdftitle={Search for stealth supersymmetry in events with jets, either photons or leptons, and low
  missing transverse momentum in pp collisions at 8 TeV},%
pdfsubject={CMS},%
pdfkeywords={CMS, physics, stealth, supersymmetry, photons, leptons, low missing transverse energy}}

\maketitle

\section{Introduction}
\label{sec:intro}
Models of supersymmetry~\cite{susy, Martin:1997ns} (SUSY) with a stable, neutral, massive, weakly
interacting, lightest supersymmetric particle (LSP) have received considerable attention in recent
years because they simultaneously offer a solution to the hierarchy problem,
allow unification of the fundamental interactions, and provide a dark matter candidate. Many
searches for SUSY are based on this scenario, which predicts large missing transverse momentum
\ptvecmiss as a consequence of the undetected LSPs. Nonetheless, well-motivated models of SUSY exist
that predict small \ptvecmiss, such as models with $R$-parity violation~\cite{rpv}, gauge mediated
SUSY breaking~\cite{giudice}, compressed spectra~\cite{martin1, martin2}, or hidden
valleys~\cite{strassler}. Many non-SUSY models of new physics, including theories with extra
dimensions~\cite{appel}, heavy-flavor compositeness~\cite{gregoire}, or little Higgs
scenarios~\cite{AH, katz}, similarly predict low-\ptvecmiss final states. As the parameter space
available for high-\ptvecmiss signatures becomes constrained by results from the CERN
LHC~\cite{Aad:2014wea, Aad:2013wta, Aad:2014mra, Aad:2015mia, Aad:2012zza, Chatrchyan:2013lya,
Chatrchyan:2014lfa, Chatrchyan:2013wxa, Chatrchyan:2012te, Chatrchyan:2012bba}, searches for these
low-\ptvecmiss alternatives become increasingly pertinent.

Among models of SUSY with low \ptvecmiss final states, the so-called stealth
scenario~\cite{stealthSusy1, stealthSusy2} has received relatively little attention. The simplest
stealth SUSY models assume low-scale SUSY breaking and introduce a new hidden sector of particles at
the weak scale, analogous to the SUSY-breaking sector, which experiences only minimal SUSY breaking
through the interactions with SM fields. Because it is weakly connected to the SUSY-breaking sector,
the hidden sector is populated with nearly mass-degenerate superpartners. With this addition, the
LSP of non-stealth scenarios, taken to be a gaugino (\ie, a neutralino or chargino), assumes the
role of the lightest ``visible sector'' SUSY particle (LVSP) and can decay without violating
$R$-parity~\cite{Farrar:1978xj} to yield a lighter hidden-sector SUSY particle. The LSP in this
model is produced from the decay of the hidden-sector SUSY particle to its SM partner, and the near
mass degeneracy of the superpartners results in the LSP being produced with low momentum. Thus,
stealth SUSY models naturally produce low-\ptvecmiss signatures with neither $R$-parity violation
nor a special tuning of masses.

\begin{figure}[htbp]
  \centering
  \includegraphics[width=0.5\textwidth]{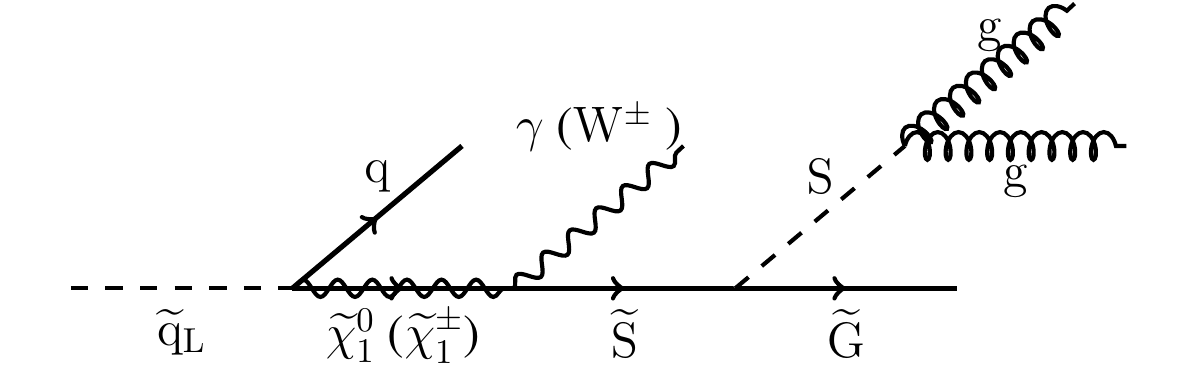}
  \caption{Decay of a squark \PSQ to a quark and gaugino \ggo in stealth SUSY. The subsequent decay
    of the gaugino produces a singlino \sgo and a $\gamma$ or $\PW^\pm$ boson, and the singlino decays
    to two gluons and a soft gravitino \PXXSG.}
  \label{fig:feynman}
\end{figure}

In this letter we present a search for stealth SUSY signatures involving the decay of a gaugino to a
stealth-model particle and either a photon ($\gamma$ analysis) or a leptonically decaying $\PW^\pm$
boson ($\ell^\pm$ analysis). The data sample, corresponding to an integrated luminosity of
19.7\fbinv of proton-proton collisions at $\sqrt{s} = 8\TeV$, was collected with the CMS detector at
the LHC in 2012. For the interpretation of results, we assume a minimal hidden sector composed of an
$R$-parity-even scalar particle S and its superpartner, the singlino \sgo, both of which are
singlets under all SM interactions. We consider singlino production in the context of squark pair
production, with the decay of the squark shown in Fig.~\ref{fig:feynman}. In the $\gamma$
($\ell^\pm$) scenario, the LVSP neutralino (chargino) decays to an \sgo and a photon ($\PW^\pm$
boson), with a subsequent decay of the \sgo to an S and a gravitino, $\sgo \to \PXXSG
\mathrm{S}$. The S is assumed to decay to jets via $\mathrm{S} \to \Pg\Pg$. Because of
the small mass splitting between the S and \sgo, the resulting gravitino carries very little
momentum and yields low \ptvecmiss.

The $\gamma$ analysis is an extension of a similar study~\cite{stealth_2012} performed with a sample
of proton-proton collisions at $\sqrt{s} = 7\TeV$. The $\ell^\pm$ analysis is the first of its kind.
For the $\gamma$ analysis we require the presence of two photons in the final state, while for the
$\ell^\pm$ analysis we require the presence of two leptons with different flavors and opposite
charges ($\Pe^\pm \mu^\mp$). Both the $\gamma$ and $\ell^\pm$ analyses are based on a search for an
excess of events with a large number of jets \nj and high \st, where \st is the scalar sum of the
transverse momenta \pt of all physics objects used in the study. We perform a statistical test for
the presence of the specific stealth SUSY models described in this letter, and provide additional
information to allow alternative interpretations of the data.

This letter is organized as follows: in Section~\ref{sec:sele} we describe the data samples, trigger
criteria, and object definitions used in the analysis. The details of the simulation of the signal
and background samples are described in Section~\ref{sec:mc}. Methods based on control samples in
data for estimating the backgrounds for the $\gamma$ and $\ell^\pm$ analyses are given in
Sections~\ref{sec:bkg_photon} and~\ref{sec:bkg_lepton}. Systematic uncertainties are discussed in
Section~\ref{sec:sysUnc} and the results, including exclusion limits, are presented in
Section~\ref{sec:res}. Section~\ref{sec:sum} summarizes our conclusions.

\section{Trigger and object selection}
\label{sec:sele}

The central feature of the CMS apparatus is a superconducting solenoid of 6\unit{m} inner diameter that
surrounds a silicon pixel and strip tracker, covering the pseudorapidity region $\abs{\eta} <
2.5$, as well as a lead tungstate crystal electromagnetic calorimeter (ECAL) and a
brass/scintillator hadron calorimeter (HCAL), both covering $\abs{\eta} < 3.0$. Muons are
detected with gas-ionization detectors embedded in the steel flux-return yoke covering the range
$\abs{\eta} < 2.4$. A more detailed description of the CMS detector, together with a definition
of the coordinate system used and the relevant kinematic variables, can be found in
Ref.~\cite{cmsdet}.

For the $\gamma$ analysis we employ a diphoton trigger requiring two photons satisfying $\pt > 36$
and 22\GeV. The SM background is studied with events from a trigger that requires $\htt >
750\GeV$, where \htt is the scalar sum of the \pt of all jets in the event with $\pt > 40\GeV$. The
$\ell^\pm$ analysis is based on a single-muon trigger, which requires the presence of at least one
muon with $\pt > 24\GeV$ and $\abs{\eta} < 2.1$.

Muon candidates are reconstructed with the particle-flow (PF) algorithm \cite{CMS:2010byl}, which
simultaneously reconstructs all particles produced in a collision based on information from all
detector subsystems and identifies each as a charged or neutral hadron, photon, muon, or electron.
Candidates are required to have $\pt > 15\GeV$, to be reconstructed in the fiducial volume of the
trigger ($\abs{\eta} < 2.1$), and to have a transverse\,(longitudinal) impact parameter less than
2\,(5)\mm with respect to the primary vertex of the event. The primary vertex is defined as the
vertex with the highest sum of $\pt^2$ of tracks associated with it. To ensure a precise measurement
of the transverse impact parameter of the muon track relative to the beam spot, we consider only
muons with tracks containing more than ten measured points in the silicon tracker and at least one
in the pixel detector. We ensure isolation from other activity in the event by restricting the
scalar \pt sum of all PF-reconstructed photons and charged and neutral hadrons within a cone $\Delta
R \equiv \sqrt{\smash[b]{(\Delta\eta)^{2} + (\Delta\phi)^{2}}} = 0.4$ around the muon direction to be less than
12\% of the candidate \pt after subtracting the contributions of additional $\Pp\Pp$ collisions
(pileup)~\cite{2012JInst7P0002T}.

Electron candidates are reconstructed by matching an energy cluster in the ECAL barrel ($\abs{\eta}
< 1.44$) with a track reconstructed with a Gaussian sum filter~\cite{CMS:2010bta} in the tracking
system. The ECAL endcap regions are omitted due to the low expected signal acceptance in these
regions. The shape of the matched ECAL cluster must be consistent with that expected for electrons,
and the difference in the inverse cluster energy and the inverse track momentum must be less than
$0.05\GeV^{-1}$. The electron candidate is required to be inconsistent with the conversion of a
photon to an $\Pep\Pem$ pair in the tracker. The track for the candidate must have a longitudinal
impact parameter less than 1\mm with respect to the primary vertex and fewer than two missing hits
in the tracker. All candidates must have $\pt > 15\GeV$, and the pileup-corrected sum of the \pt of
all PF-reconstructed charged hadrons, neutral hadrons, and photons in a cone of radius $\Delta R =
0.3$ around the candidate direction is required to be less than 10\% of the candidate \pt.

Photon candidates are reconstructed from energy clusters in the ECAL barrel with $\abs{\eta} <
1.44$. We require the ECAL cluster shape to be consistent with that expected for photons, and the
energy detected in the HCAL in the direction of the photon shower not to exceed 5\% of the ECAL
energy. A base requirement of $\pt > 15\GeV$ is imposed on all photon candidates. Further, the
candidate cannot be matched to hit patterns in the pixel detector. In a cone of radius $\Delta R =
0.3$ around the candidate photon direction, the pileup-corrected charged-hadron contribution must be
less than 1.5\GeV, the corrected neutral-hadron contribution less than $1.0\GeV + 4\%$ of the photon
\pt, and the corrected electromagnetic contribution less than $0.7\GeV + 0.5\%$ of the photon \pt.

Jets are reconstructed with the anti-\kt clustering algorithm~\cite{Cacciari:2008gp} with a
distance parameter of 0.5 using PF objects as input~\cite{CMS:2009nxa}. To remove jets arising from
potential instrumental and non-collision backgrounds, we require the fraction of jet energy coming
from charged and neutral electromagnetic deposits to be less than 0.99, the neutral hadron
fraction to be less than 0.99, and the charged hadron fraction to be greater than zero. The jet
energy and momentum are corrected for the nonlinear response of the calorimeter and the effects of
pileup~\cite{Chatrchyan:2011ds}.
Jets are required to have corrected $\pt > 30\GeV$, $\abs{\eta} < 2.4$, and to be isolated from
photon and lepton candidates by $\Delta R > 0.5$. Jets are identified as originating from b-quark
hadronization (b-tagged) using a combined secondary vertex algorithm that yields 70\% signal
efficiency for b jets and 1.5\% misidentification of light quark jets.~\cite{Chatrchyan:2012jua}.

The missing transverse momentum vector \ptvecmiss is defined as the projection on the plane
perpendicular to the beams of the negative vector sum of the momenta of all reconstructed particles
in an event. Its magnitude is referred to as \ETm. \st is the scalar \pt sum of all accepted physics
objects in the analysis: muons, electrons, photons, jets, and \ETm.

\section{Simulation of background and signal events}
\label{sec:mc}

Monte Carlo (MC) simulations of signal and background processes are used to optimize selection
criteria, validate analysis performance, determine signal efficiencies, and determine some
backgrounds and systematic uncertainties. To simulate these samples, we use the
\MADGRAPH 5.1.3.30~\cite{MADGRAPH} leading-order (LO) event generator unless otherwise noted. The
\PYTHIA 6.426~\cite{PYTHIA} event generator with CTEQ6L1~\cite{cteq6l1} parton distribution
functions (PDF) and parameters based on measurements from the LHC run at $\sqrt{s} = 7\TeV$ is used
to describe parton showering, hadronization, multiple-parton interactions, and the underlying event
for \MADGRAPH5 samples.
A full simulation of the CMS detector based on the \GEANTfour~\cite{GEANT4} package is applied to
all samples. Each event is superimposed with a set of simulated minimum bias events to reproduce the
effect of pileup.

For the $\gamma$ analysis, SM diphoton events are generated by requiring exactly two photons with
$\pt > 20\GeV$ and minimum separation $\Delta R = 0.4$. Up to four additional partons are allowed.
For the $\ell^\pm$ analysis, we generate samples of events with a top quark-antiquark (\ttbar) pair,
Drell--Yan (DY), $\cPZ\cPZ$, $\PW\PW$, and $\PW\cPZ$ production. The \ttbar sample is produced with up to three
additional partons, the DY sample is produced with up to four additional partons, and the diboson
samples are produced with up to two additional partons. Single-top quark ($t$-, $s$-, and
t\PW-channels) samples are generated with the \POWHEG v1.0~\cite{Alioli:2009je, Re:2010bp,
Nason:2004rx, Frixione:2007vw, Alioli:2010xd} generator. The \ttbar and DY samples are normalized to
cross sections calculated at next-to-next-to-leading-order accuracy~\cite{topnnlo, dynnlo}. The
normalizations of the single-top quark and diboson samples are valid to next-to-leading-order
(NLO)~\cite{singletopnnlo} and LO~\cite{Campbell:2011bn}, respectively. The diphoton sample is used only to validate the background estimation method and so its normalization is not relevant.

We generate signal samples for both analyses using the \PYTHIA generator with the CMS fast
simulation~\cite{Abdullin:2011zz} of the detector. The models are characterized by the masses of the
particles in the decay chain. The small \sgo-S mass splitting, the central feature of stealth SUSY,
is taken to be 10\GeV, and we assume the \sgo mass to be 100\GeV. In the $\ell^\pm$ analysis, a
range of squark masses ($M_{\PSQ}$) are considered from 300 to 1000\GeV, and the chargino is fixed
to be half of $M_{\PSQ}$ rounded up to the nearest 100\GeV. In the $\gamma$ analysis, $M_{\PSQ}$
ranges from 200 to 1400\GeV and the neutralino mass ($M_{\ggo}$) ranges from 150 to 1350\GeV, with
the requirement $M_{\ggo} < M_{\PSQ}$. In both models, the gravitino mass is taken to be zero. We
assume branching fractions of unity for the decays $\PSGcz_{1} \to \sgo \gamma$ and
$\PSGcpm_{1} \to \sgo \PW^{\pm}$ in the $\gamma$ and $\ell^\pm$ analyses, respectively.

The production cross sections for these processes are calculated as a function of $M_{\PSQ}$ at NLO
accuracy including the resummation of soft gluon emission at next-to-leading logarithmic (NLL)
accuracy~\cite{Beenakker:1996ch, Kulesza:2008jb, Kulesza:2009kq, Beenakker:2009ha} with
uncertainties computed as described in Ref.~\cite{nloxsec}. The $\PSQ \to \Pq \PSGcpm_{1}$ decay is
possible only for left-handed squarks, so for consistency the production processes are limited to
$s$-channel production of mass-degenerate, left-handed squarks ($\PSQu$, $\PSQd$, $\PSQs$, and
$\PSQc$) for both analyses. The masses of the gluino, the right-handed squarks, and top and bottom
squarks are assumed to be too large to participate in the interactions. The masses of the gluino and
right-handed squarks have been changed with respect to the previous analysis~\cite{stealth_2012},
where they were assumed to be sufficiently light to participate in the production.

\section{The \texorpdfstring{$\gamma$}{gamma} analysis}
\label{sec:bkg_photon}

The dominant backgrounds for the $\gamma$ analysis arise from the SM production of events with two
photons, and with a photon and a jet misidentified as a photon. We estimate these backgrounds as
functions of \st and \nj directly from the data via the \st shape invariance method~\cite{cmsBH2010,
Chatrchyan:2012taa, bh3, stealth_2012}, which relies on the empirical observation that the shape of
the \st distribution is independent of the number of jets in the final state above some \st
threshold. Thus, the \st shape obtained from a low-\nj control sample can be used to predict the
background in the high-\nj signal sample. This method is validated with a data control sample and
simulation.

Starting from the basic object selection described in Section~\ref{sec:sele}, the $\gamma$ analysis
imposes two sets of selection criteria based on the trigger used to collect the data, as indicated
in Table~\ref{tab:selections}. Selection A, which is applied to the diphoton simulation and to
events in the data that satisfy the diphoton trigger, requires a photon with $\pt > 40\GeV$, a
second photon with $\pt > 25\GeV$, and at least two jets. Selection B is applied to events passing
the $\htt$ trigger and requires $\htt > 800\GeV$, exactly one photon with $\pt > 15\GeV$, and at
least two jets. Additionally, we require $\pt < 75\GeV$ for the photon to make this sample disjoint
from a single photon selection, not discussed here, that was used to test the background estimation
method. Events that satisfy selection B, along with simulated diphoton events, are used to validate
the background estimation method. Events that satisfy selection A are further divided into three
samples, shown in Table~\ref{tab:bkgSidebands}: the signal-enhanced ``search region'' is defined as
events with $\nj \geq 4$ and $\st > 1200\GeV$, the signal-depleted ``\st sideband'' is defined as
events with $\nj \geq 4$ and $1100 < \st < 1200\GeV$, and the signal-depleted ``\nj sideband'' is
defined as events with $\nj = 3$ and $\st > 1100\GeV$.

\begin{table}
  \centering
    \topcaption{Selection criteria for the search (A) and control (B) regions for the $\gamma$
      analysis based on the \pt of the photons and the \htt in the event.}
    \begin{tabular}{lcccr}
      Selection & \nj & $\gamma_{1}$ \pt & $\gamma_{2}$ \pt & \multicolumn{1}{c}{\htt} \\
      &(\GeVns{})&(\GeVns{})&\multicolumn{1}{c}{(\GeVns{})}\\
      \hline
      A & $\geq$2 & $>$40 & $>$25 & $>$60  \\
      B & $\geq$2 & $<$75 & ---    & $>$800 \\
    \end{tabular}
    \label{tab:selections}

\end{table}

\begin{table}
  \centering
    \topcaption{Selection criteria defining the search and sideband regions for events passing
      selection A for the $\gamma$ analysis based on the number of jets and the \st in the event.}
    \begin{tabular}{lcc}
      Region        & \nj      & \st\,(\GeVns{})  \\
      \hline
      Search        & $\geq$4 & $>$1200   \\
      \st sideband  & $\geq$4 & 1100--1200 \\
      \nj sideband  & $=$3    & $>$1100   \\
    \end{tabular}
    \label{tab:bkgSidebands}

\end{table}

\begin{figure}[htbp]
  \centering
  \includegraphics[width=0.49\textwidth]{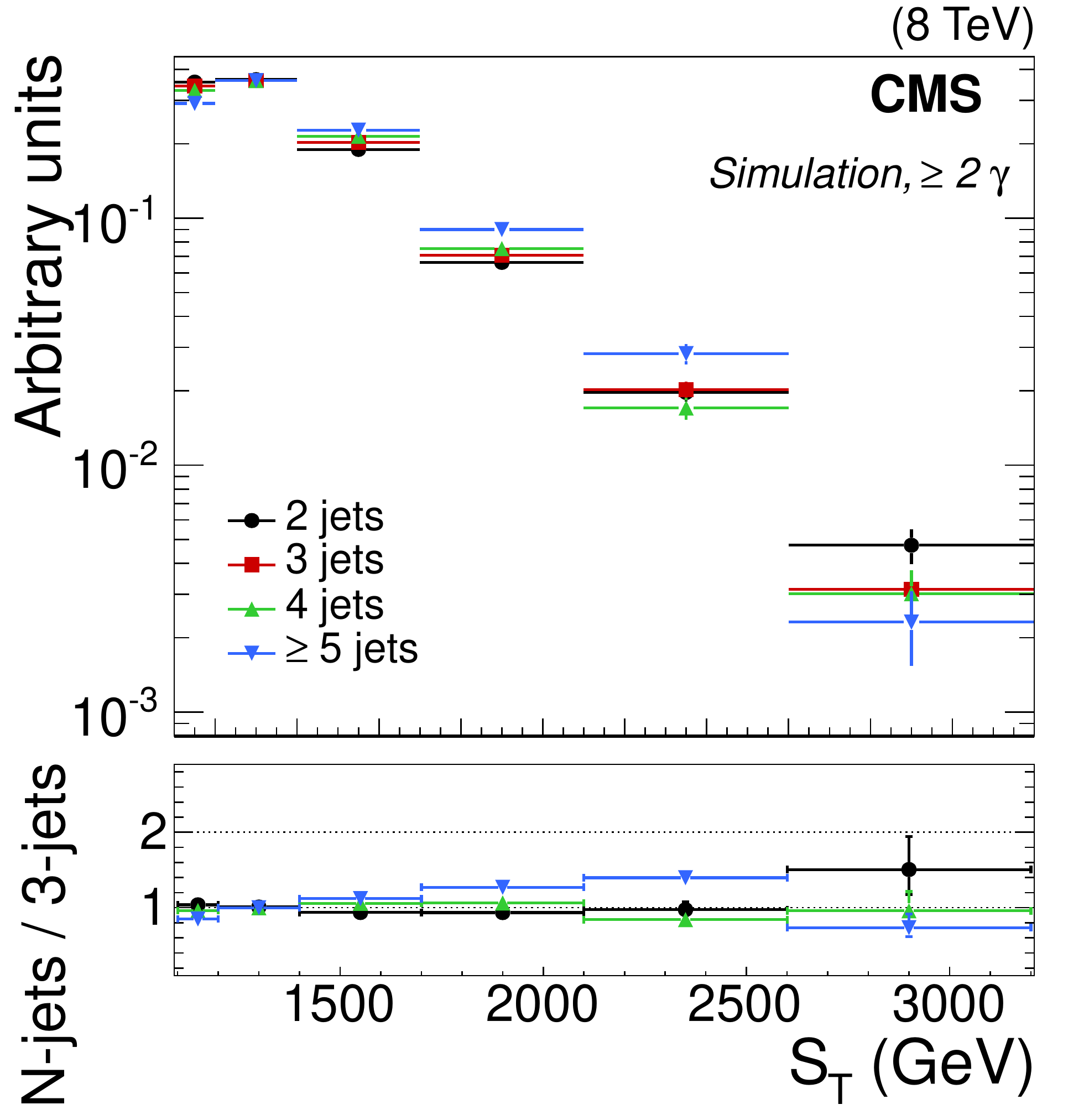}
  \includegraphics[width=0.49\textwidth]{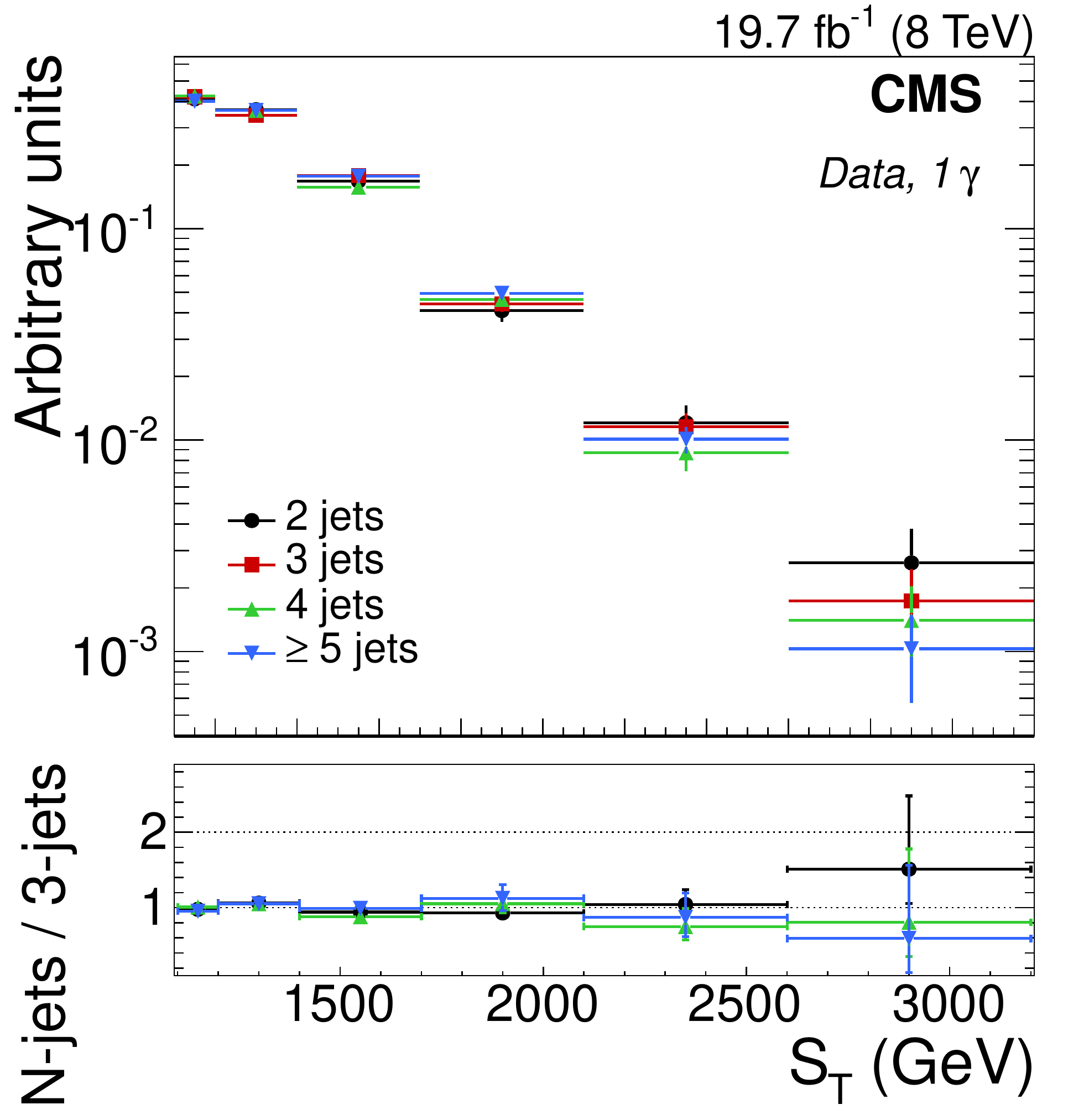}
  \caption{\st distributions used in the $\gamma$ analysis as a function of \nj for simulated
    diphoton events passing selection A (\cmsLeft) and for data events passing selection B (\cmsRight). The
    distributions are normalized to unit area. The lower plots show ratios with respect to the $\nj
    = 3$ distribution.}
  \label{fig:st}
\end{figure}

To verify the assumption that the \st distribution is independent of \nj, we present in
Fig.~\ref{fig:st} the \st spectra for events with 2, 3, 4, and $\geq$5 jets. The assumption is
checked in simulated events passing selection A (\cmsLeft) and directly in data for events passing
selection B (\cmsRight).
The distributions are normalized to unit area and the lower plots show their ratios with respect to
the $\nj = 3$ distribution. For the selection B data, the ratios are seen to be consistent with a
constant function of \st within the uncertainties. For the simulated diphoton sample, the
$\nj \geq 5$ events show an upward trend with increasing \st with respect to the $\nj = 3$
distribution. The increase corresponds to a 15\% increase in the expected background rate for
$\st > 1200\GeV$ and is accounted for in the evaluation of systematic uncertainties, as described in
Section~\ref{sec:sysUnc}.

To obtain the shape of the \st distribution for the SM background in the search region, we fit the
\st distribution in the \nj sideband with the nominal shape $1 / x^{p_{1}\ln\st}$, where $x \equiv
\st/(8000\GeV)$. Two alternate functions, $1 / x^{p_{2}}$ and $1 / e^{p_{3}x}$, are used to assess
the systematic uncertainty associated with the choice of fit function. We find $p_{1} = 1.01 \pm
0.19$. The normalization of this shape is obtained from events in the \st sideband.

\section{The \texorpdfstring{$\ell^\pm$}{ell+/-} analysis}
\label{sec:bkg_lepton}

For the $\ell^\pm$ analysis, the signal region is defined in terms of \nj, the number \nbj of
b-tagged jets, and the lepton flavors and charges. To reduce the multijet and $\PW$+jets backgrounds,
we require that both $\PW$ bosons decay leptonically resulting in exactly two oppositely charged
leptons in the final state with no additional lepton that satisfies loosened isolation criteria. To
reduce the large DY contribution to the background, we require one of these leptons to be a muon and
the other to be an electron. To ensure optimal trigger efficiency, the muon is required to have $\pt
> 30\GeV$. Finally, to suppress the \ttbar background, signal events are required to have $\nbj =
0$. The principal requirements for the signal event selection are listed in the top row of
Table~\ref{tab:samples}. To enhance the statistical significance of a potential observation, we
divide the signal sample into four exclusive regions based on \nj (4, 5, 6, and $\geq$7) and divide
each \nj bin into three inclusive samples with \st thresholds of 300, 700, and 1200\GeV. These
threshold values were determined through a procedure that optimizes sensitivity to stealth SUSY
production via examination of the $\cPZ_\mathrm{Bi}$ variable~\cite{zbi}, which is the ratio of the
Poisson means of the expected signal and background given the systematic uncertainty in the expected
background. We find that thresholds of $\st^\text{min} =$ 300, 700, 700, and 1200\GeV are optimal for
squark mass values of 300, 400, 500, and 600\GeV, respectively.

The largest SM background contributions in the signal regions are from \ttbar and single-top quark
events, which we collectively refer to as the ``top-quark background''. Depending on the \st
threshold, approximately 1--10\% of the background arises from $\cPZ \to \tau^{+}\tau^{-}$,
diboson, and non-prompt lepton production, where ``non-prompt'' refers to leptons from hadron decay
and to hadrons that are misidentified as leptons. The estimate of the SM background is based on four
data control regions, defined in the bottom four rows of Table~\ref{tab:samples} in terms of \nj,
\nbj, and the lepton flavors and charges.

\begin{table}
  \centering
    \topcaption{Summary of search and control sample definitions for the $\ell^\pm$ analysis based on
      the number of jets, number of b-tagged jets, lepton flavor, and lepton charge.}
    \begin{tabular}{lcrr}
      Sample            & Leptons              & \multicolumn{1}{c}{\nj}      & \multicolumn{1}{c}{\nbj}     \\
      \hline
      Search            & $\Pe^\pm$, $\mu^\mp$   & $\geq$4 & 0        \\
      Top shape         & $\Pe^\pm$, $\mu^\mp$   & $\geq$2 & $\geq$2 \\
      Top normalization & $\Pe^\pm$, $\mu^\mp$   & $<$4    & 0        \\
      Drell--Yan        & $\mu^\pm$, $\mu^\mp$ & $\geq$2 & 0        \\
      Non-Prompt        & $\Pe^\pm$, $\mu^\pm$   & $\geq$2 & 0        \\
    \end{tabular}
    \label{tab:samples}

\end{table}

\begin{figure}[htbp]
  \centering
  \includegraphics[width=0.5\textwidth]{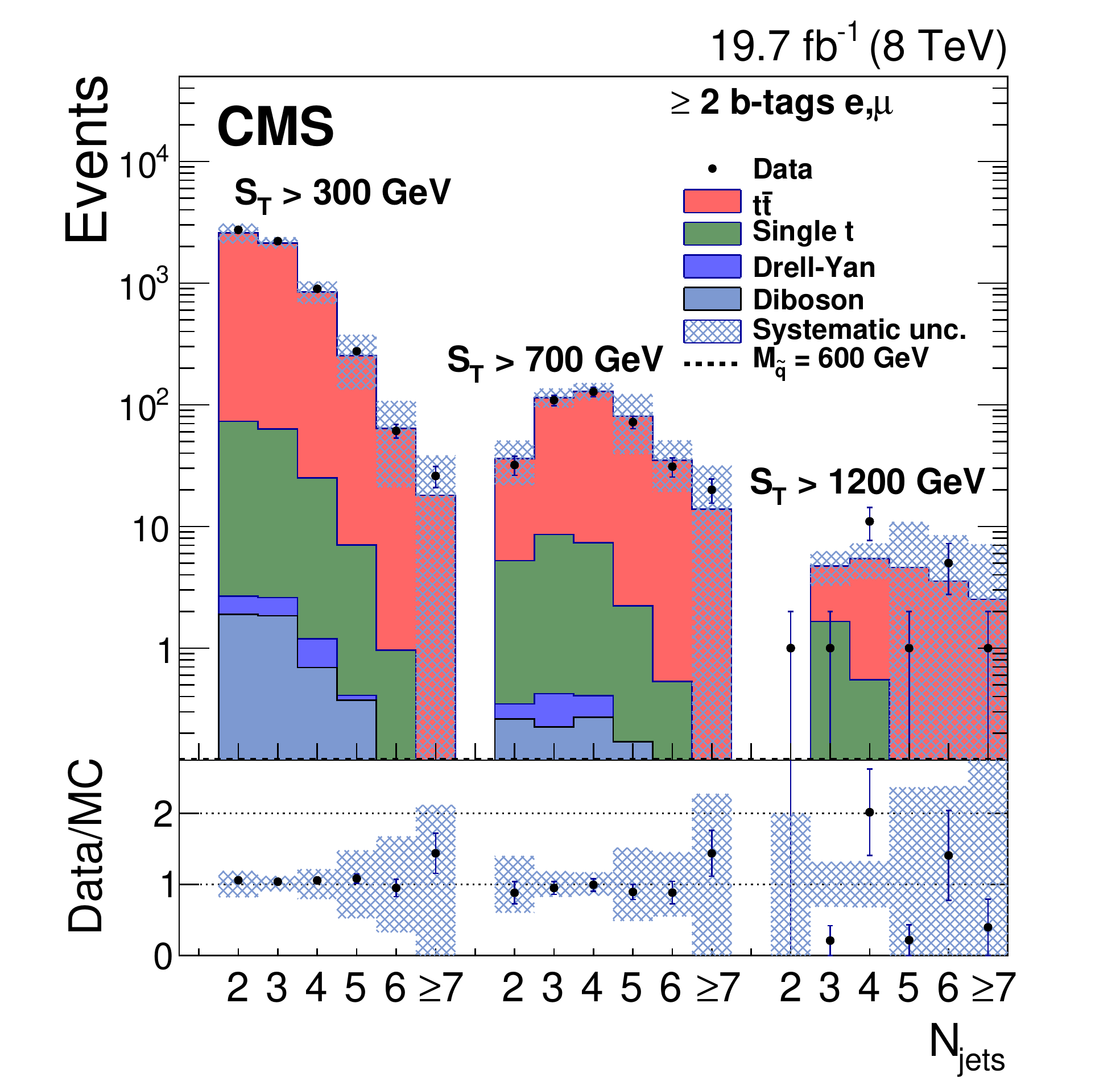}
  \caption{Distribution of \nj for data and simulation, for the top-shape control region used in the
    $\ell^\pm$ analysis. The lower plot shows the ratio of data and simulation, with systematic
    uncertainties shown by the shaded bands. The (negligible) signal contribution to this control
    sample is shown as a dashed line that appears to coincide with the horizontal axis.}
  \label{fig:eventCategories}
\end{figure}

The top-quark background is estimated from simulation, with corrections to the shape of the \nj
distribution obtained by comparing data and simulation in the ``top shape'' control region defined
in Table~\ref{tab:samples}. A comparison of data and simulation in this control region is shown in
Fig.~\ref{fig:eventCategories} with the systematic uncertainty in the top quark background,
estimated by varying the renormalization and factorization scale up and down by a factor of 2. The
small corrections, which are derived from the lowest \st bin, are consistent with unity for all
values of \nj. The top-quark simulation is then normalized to the data in the ``top normalization''
control region defined in Table~\ref{tab:samples}. Before obtaining the normalization correction
factor from this sample, we use the simulation to subtract contributions from the DY, diboson, and
non-prompt backgrounds, which collectively account for 20\% of the total background. We then
determine the correction factor from events with $\st > 200\GeV$ as the ratio of the number of
events in this background-subtracted data sample to the number of events in the simulated top-quark
background, finding $0.97 \pm 0.02$, where the uncertainty is statistical.

Similarly, the small DY background (about 10\% of the total background) is evaluated from
simulation, with a correction factor for the normalization derived from the DY control sample
(Table~\ref{tab:samples}), which requires two oppositely charged muons. Because the contribution of
signal events to the DY control sample is potentially significant at large \nj, we perform a fit to
the dimuon mass spectrum using templates from simulation to describe the shapes of the DY and
diboson components, with a first-order polynomial used to describe the combined shape of potential
signal and remaining (non-peaking) background events. The number of DY events $N_\mathrm{DY}$,
polynomial normalization, and polynomial slope are determined in the fit. The correction factor,
defined as the ratio of $N_\mathrm{DY}$ to the number of events in the DY simulation, ranges from $1.02
\pm 0.01$ for $\nj = 2$ to $1.56 \pm 0.25$ for $\nj \geq 6$, where the uncertainties are
statistical.

To estimate the small background associated with non-prompt leptons (about 2\% of the total
background) we use the non-prompt control sample (Table~\ref{tab:samples}), defined using same
charge (SC) $\Pe\mu$ events. After subtracting the simulated contribution to this sample from SM
top-quark and diboson events, we take the remaining data as the estimate of the non-prompt
background in the search region. Because of the low number of SC events with high \nj and high \st,
we fit the \nj distribution to an exponential function for $\st > 300\GeV$. The normalization of the
exponential distribution is determined for each \st threshold by the total number of events passing
the selection.

To estimate the diboson background (about 10\% of the total background) we use the prediction from
simulation.

\section{Systematic uncertainties}
\label{sec:sysUnc}

We evaluate the systematic uncertainties in the background expectation, signal efficiency, and
luminosity. For each source of uncertainty, we describe below the uncertainty value and the method
used for its estimation.

For the $\gamma$ analysis, the largest systematic uncertainty in the background prediction arises
from the statistical uncertainty in the normalization of the background shape from the \st sideband,
which is $30\%$ ($38\%$) for $\nj = 4$ ($\geq$5). The largest uncertainty in the assumed shape of
the \st distribution is due to the statistical uncertainty in the estimation of the fitted parameter
$p_1$ (Section~\ref{sec:bkg_photon}), which results in a systematic uncertainty of 31\% for $\st >
1200\GeV$. The second largest uncertainty associated with the shape arises from the assumption that
the \st shape is independent of \nj. We estimate this uncertainty by first separately fitting the
\st distributions for $\nj = 4$ and $\nj \geq 5$ to the nominal function, for the diphoton
simulation in the selection A region and for the data in the selection B region. We then compare the
resulting fitted parameter values with the nominal results for $\nj = 3$ in the corresponding sample
and take the largest difference as the systematic uncertainty in the values of the parameters. The
largest difference is observed for $\nj \geq 5$ and corresponds to a systematic uncertainty of 15\%
in the background prediction. The smallest shape uncertainty, which is related to the choice of the
fit function, is evaluated by constructing the envelope formed by the nominal fit function and the
two alternate fit functions described in Section~\ref{sec:bkg_photon} and results in a 12\%
variation in the total background prediction for $\st > 1200\GeV$.

The dominant systematic uncertainty in the $\ell^\pm$ analysis is associated with the top-quark
background. The uncertainty in the \nj shape corrections for the top-quark background is dominated
by the statistical uncertainty in the control sample and is estimated to be 2--25\% depending on
\nj. The uncertainty in the normalization is determined by finding the correction as described in
Section~\ref{sec:bkg_lepton} for $300 < \st < 700\GeV$ and $\st > 700\GeV$ separately.  We find
corrections of $0.97 \pm 0.02$ and $0.86 \pm 0.12$ respectively, and take the difference summed in
quadrature with the statistical uncertainty as the systematic term, which results in a systematic
uncertainty of 15\% in the background prediction. An additional uncertainty is obtained by
simultaneously changing the renormalization and factorization scales in the simulation  by a factor
of 2 and by a factor of 0.5, resulting in a 10\% systematic uncertainty in the background
prediction. We vary the b-tagging efficiency and misidentification rates by their
uncertainties~\cite{CMS:2013vea} and find that the effect on the top background prediction varies by
1--3\% depending on \nj.

For the DY background, the uncertainty is taken to be half of the correction applied to the
simulation, and constitutes a 2--28\% uncertainty depending on \nj. For the diboson prediction the
uncertainty is given by the sum in quadrature of the difference between the CMS
measurement~\cite{cmsdiboson} and the NLO calculation of the $\PWp\PWm$ cross
section~\cite{Campbell:2011bn} and the \nj-dependent DY uncertainty. Finally, the uncertainty in the
non-prompt dilepton background comes from the statistical uncertainty in the control sample and is
50--120\% depending on the \st threshold.

The signal efficiency uncertainties for the $\gamma$ analysis are related to the statistical
uncertainty from the finite size of signal simulation samples (2--15\%, depending on \nj), knowledge
of the jet energy scale (1--7\%, depending on the \PSQ-\ggo mass difference), and photon
identification and reconstruction efficiencies (3\%). For the $\ell^\pm$ analysis, the uncertainty
due to the jet energy scale is 5\%. We assign an uncertainty of 1\% to account for the muon trigger
and reconstruction efficiencies, 3\% to account for the electron reconstruction efficiency, and
0--7\% (depending on the \st threshold and \nj) to account for the finite size of the simulated
event samples. For both analyses the uncertainty related to the size of the data sample is
2.6\%~\cite{lumi}, while the uncertainties related to the PDFs and pileup interactions are found to
be negligible.

\section{Results}
\label{sec:res}

\begin{figure}[htbp]
  \centering
  \includegraphics[width=0.49\textwidth]{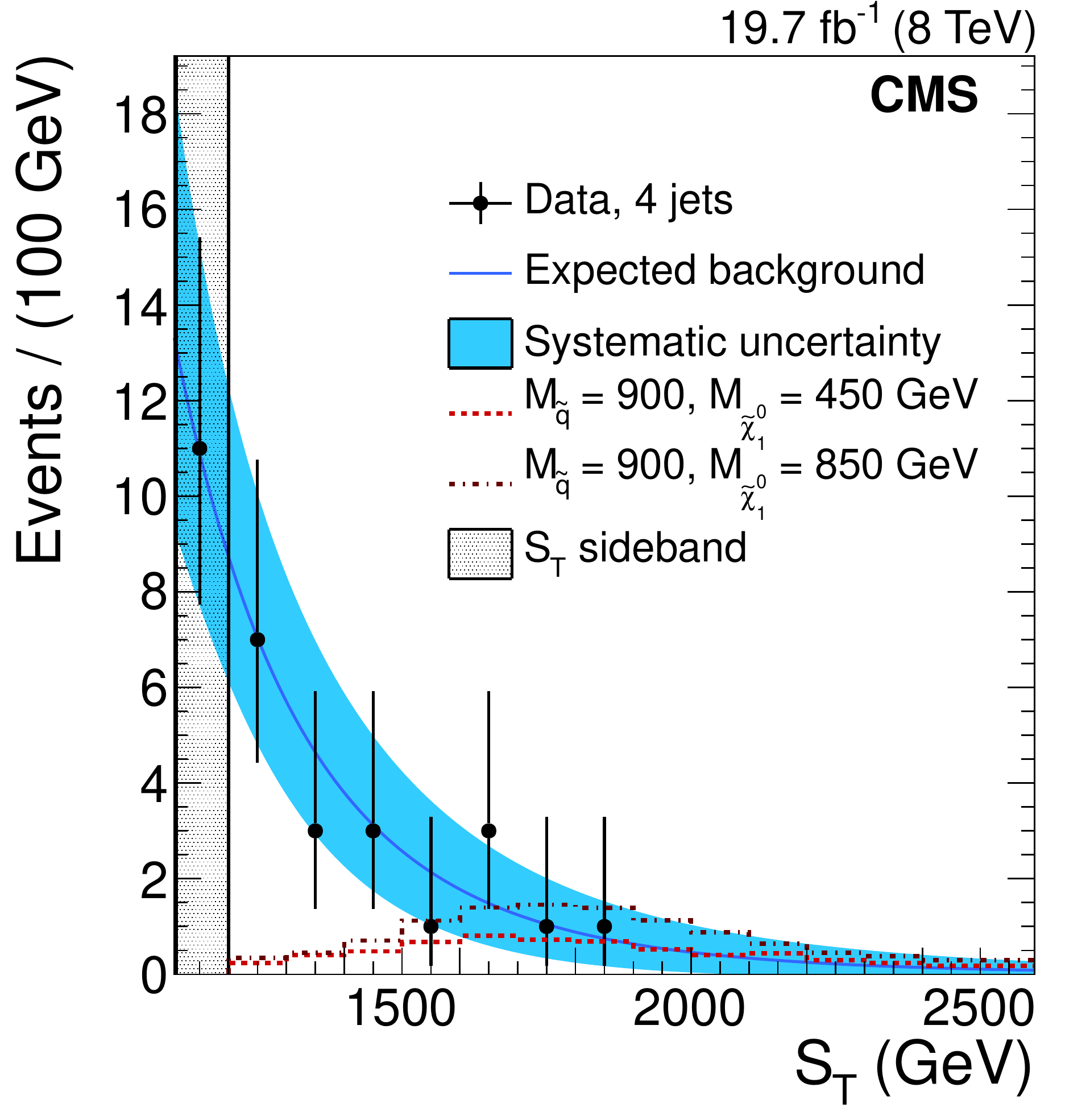}
  \includegraphics[width=0.49\textwidth]{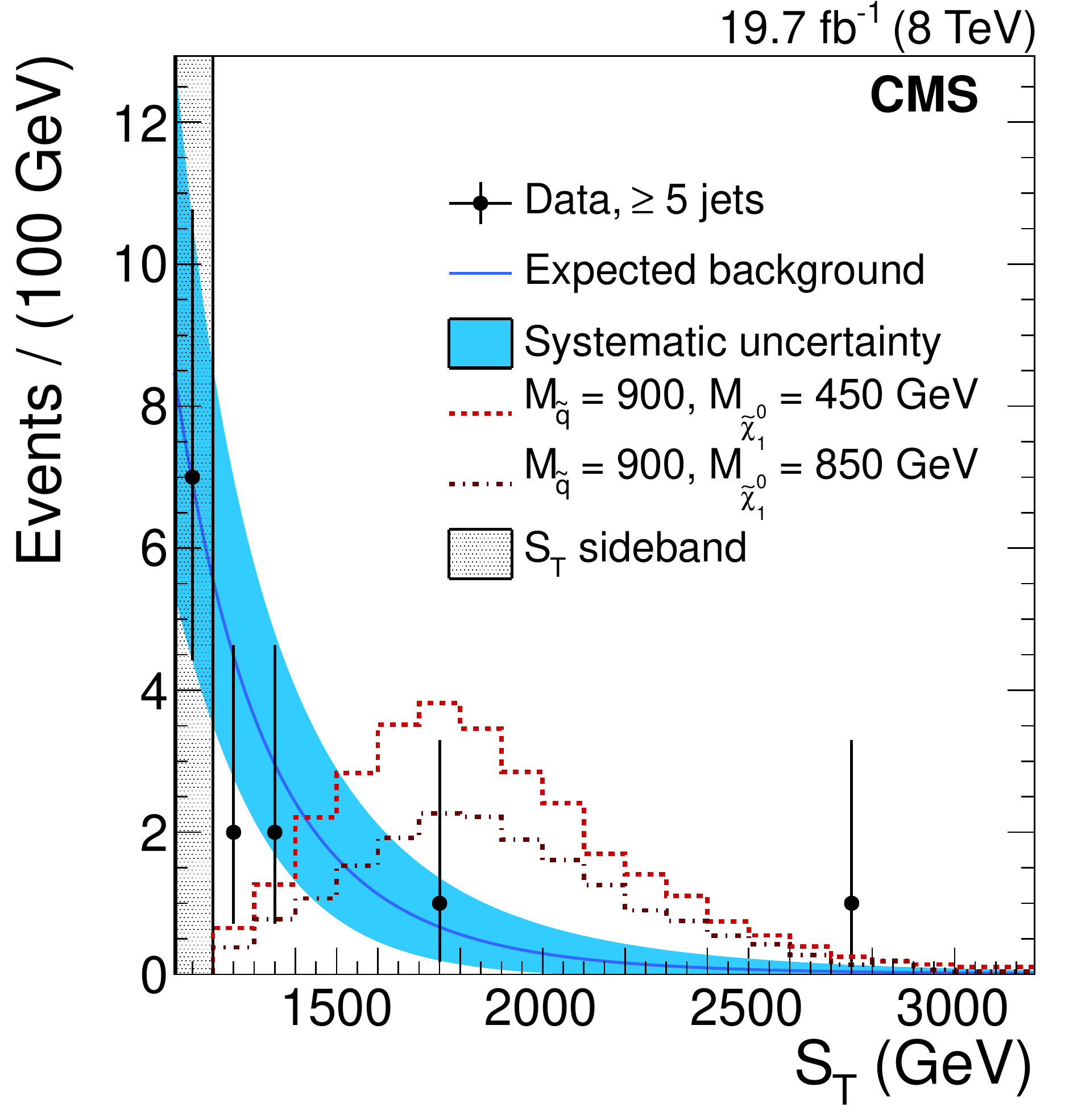}
  \caption{Measured \st distribution in comparison with the background prediction in the signal
    region of the $\gamma$ analysis for $\nj = 4$ (\cmsLeft) and $\nj \geq 5$ (\cmsRight). The systematic
    uncertainty of the background prediction and the expected distribution of signal events for
    $M_{\PSQ} = 900\GeV$ and either $M_{\ggo^{0}} = 450$ or 850\GeV are also shown.}
  \label{fig:result_doublePhoton}
\end{figure}

For the $\gamma$ analysis, the measured \st distribution and corresponding background predictions
are shown in Fig.~\ref{fig:result_doublePhoton}. We observe 19\,(6) events for $\nj = 4$\,($\geq$
5), compared to an expected background of $22.5 \pm 11.5$\,($14.3 \pm 8.1$) events. The data are
seen to agree with the background estimate within the uncertainties.

\begin{figure}[htbp]
  \centering
  \includegraphics[width=0.49\textwidth]{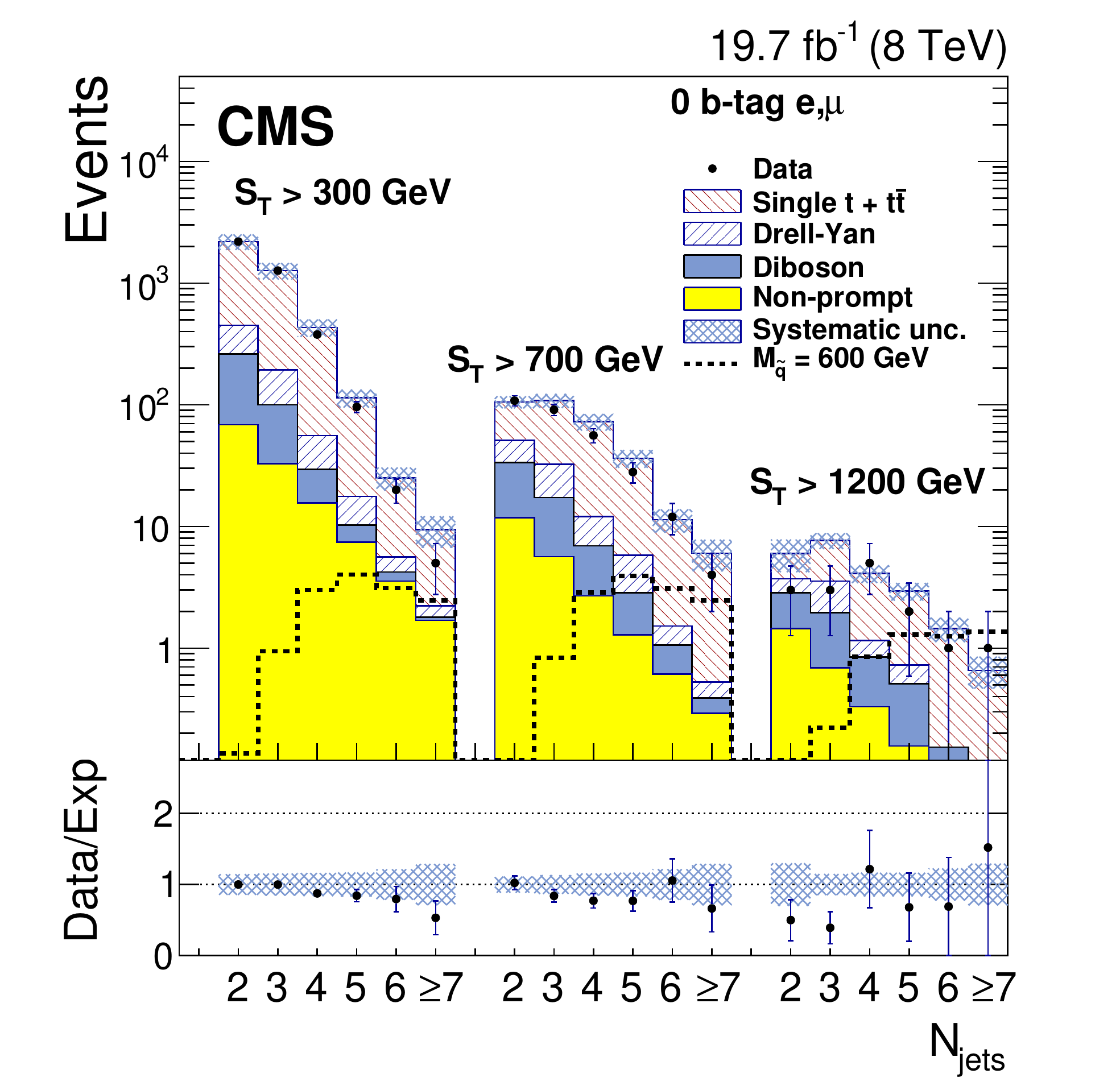}
  \caption{Measured \nj distributions in comparison with the background predictions in the signal
    regions of the $\ell^{\pm}$ analysis. The lower plots show the ratio of the data to the
    background prediction, with the systematic uncertainty in the background prediction derived from
    control samples in data.}
  \label{fig:result_lepton}
\end{figure}

Figure~\ref{fig:result_lepton} shows the corresponding results for the $\ell^\pm$ analysis. The
event yields for $\st > 1200\GeV$ are listed in Table~\ref{tab:lepEvtYield} with the total (stat. +
syst.) uncertainties. The data are seen to agree with the background expectations.

\begin{table*}
  \centering
    \topcaption{Event yields observed in data and the expected contributions from backgrounds in the
      search region of the $\ell^{\pm}$ analysis for $\st > 1200\GeV$. The total (stat. + syst.)
      uncertainties are also shown.}
    \begin{tabular}{lcccc}
       & $\nj = 4$ & $\nj = 5$ & $\nj = 6$ & $\nj \geq 7$ \\
      \hline
      Observed events                        &  5              &  2              &  1              &  1              \\
      Total background                       & $4.14 \pm 0.68$ & $2.95 \pm 0.48$ & $1.45 \pm 0.33$ & $0.66 \pm 0.19$ \\
      Top                                    & $2.96 \pm 0.55$ & $2.22 \pm 0.43$ & $1.30 \pm 0.30$ & $0.56 \pm 0.17$ \\
      DY                                     & $0.31 \pm 0.02$ & $0.22 \pm 0.02$ & $0.00 \pm 0.02$ & $0.00 \pm 0.02$ \\
      Diboson                                & $0.58 \pm 0.18$ & $0.36 \pm 0.12$ & $0.08 \pm 0.03$ & $0.06 \pm 0.02$ \\
      Non-prompt                             & $0.30 \pm 0.36$ & $0.15 \pm 0.18$ & $0.08 \pm 0.09$ & $0.04 \pm 0.05$ \\
      Signal ($M_{\PSQ} = 600\GeV$) & $0.9  \pm 0.1$  & $1.3  \pm 0.1$  & $1.3  \pm 0.1$  & $1.4  \pm 0.1$  \\

    \end{tabular}
    \label{tab:lepEvtYield}

\end{table*}

We determine 95\% confidence level (CL) upper limits on the squark pair production cross section in
the stealth SUSY framework described above. We use the modified frequentist CL$_\mathrm{S}$
method~\cite{0954-3899-28-10-313, Junk:1999kv} based on a log-likelihood ratio test statistic that
compares the likelihood of the SM-only hypothesis to the likelihood of the presence of signal in
addition to the SM contributions.
For the $\gamma$ analysis, the likelihood functions for $\nj = 4$ and $\nj \geq 5$ are based on the
expected shapes of the \st distributions for signal and background, and the total likelihood
function is the product of the two. For the $\ell^\pm$ analysis we perform a simultaneous comparison
of the number of signal and background events passing the optimized $\st^\text{min}$ threshold
defined in Section~\ref{sec:bkg_lepton} in the $\nj = 4$, 5, 6, and $\geq$7 samples, with the
likelihood function given by the product of Poisson likelihood terms from each of the \nj regions.

Systematic uncertainties are incorporated into the test statistic as nuisance parameters, with gamma
distributions for the probability density functions for the background normalization uncertainty in
the $\gamma$ analysis and the top-quark background normalization in the $\ell^\pm$ analysis. The
probability distributions for all other uncertainties are taken to be log-normal. For the $\gamma$
analysis, the background shape uncertainties are included with full correlations in \st. For the
$\ell^\pm$ analysis, all uncertainties except those arising from statistical uncertainties in the
control samples are taken to be correlated across the \nj bins.

\begin{figure}[htbp]
  \centering
  \includegraphics[width=0.49\textwidth]{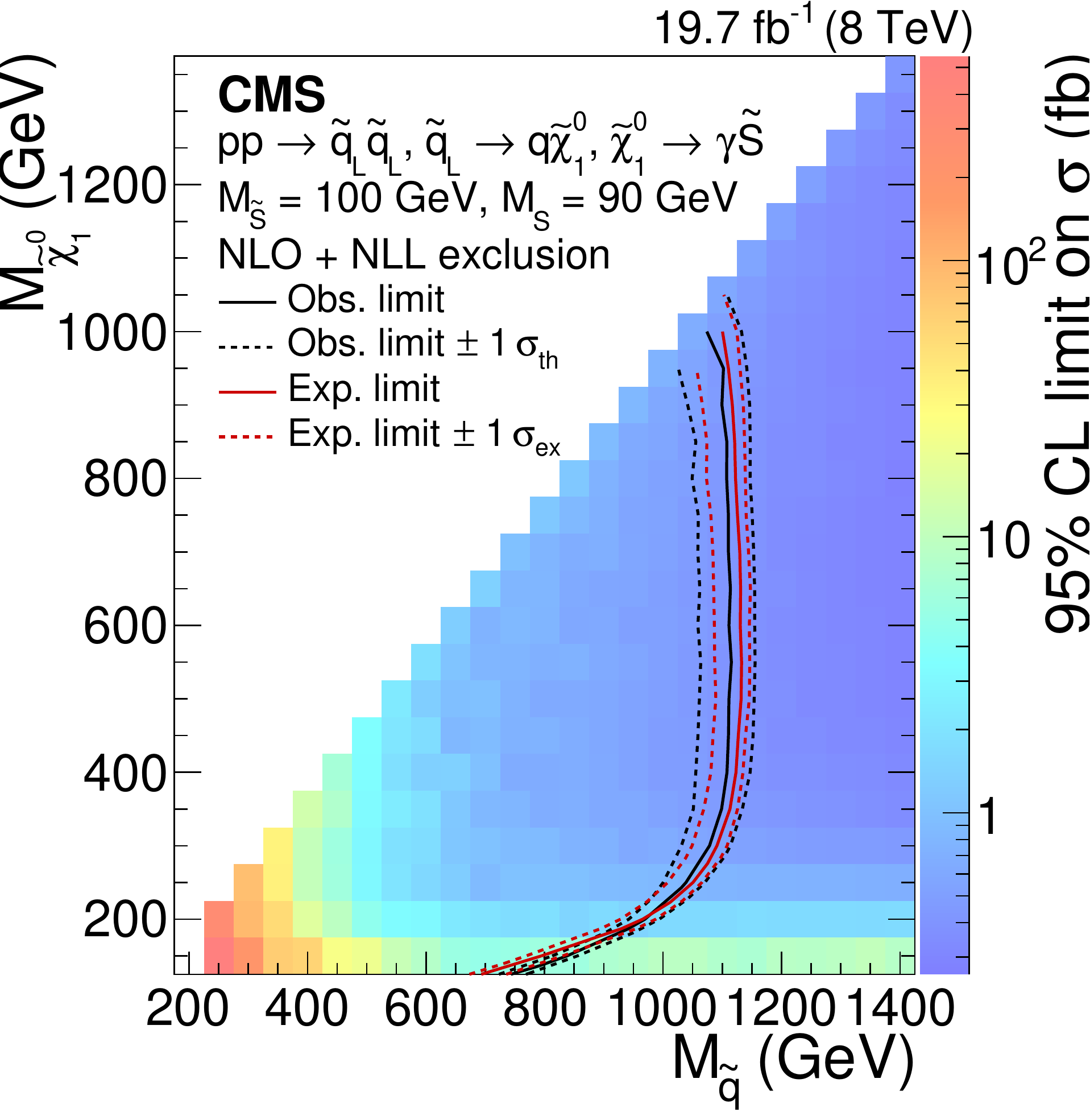}
  \caption{The 95\% confidence level upper limits on the squark pair production cross section as a
    function of squark and neutralino masses from the $\gamma$ analysis. The contours show the
    observed and median expected exclusions assuming the NLO+NLL cross sections, with their one
    standard deviation uncertainties.}
  \label{fig:totalLimit_doublePhoton}
\end{figure}

\begin{figure}[htbp]
  \centering
  \includegraphics[width=0.49\textwidth]{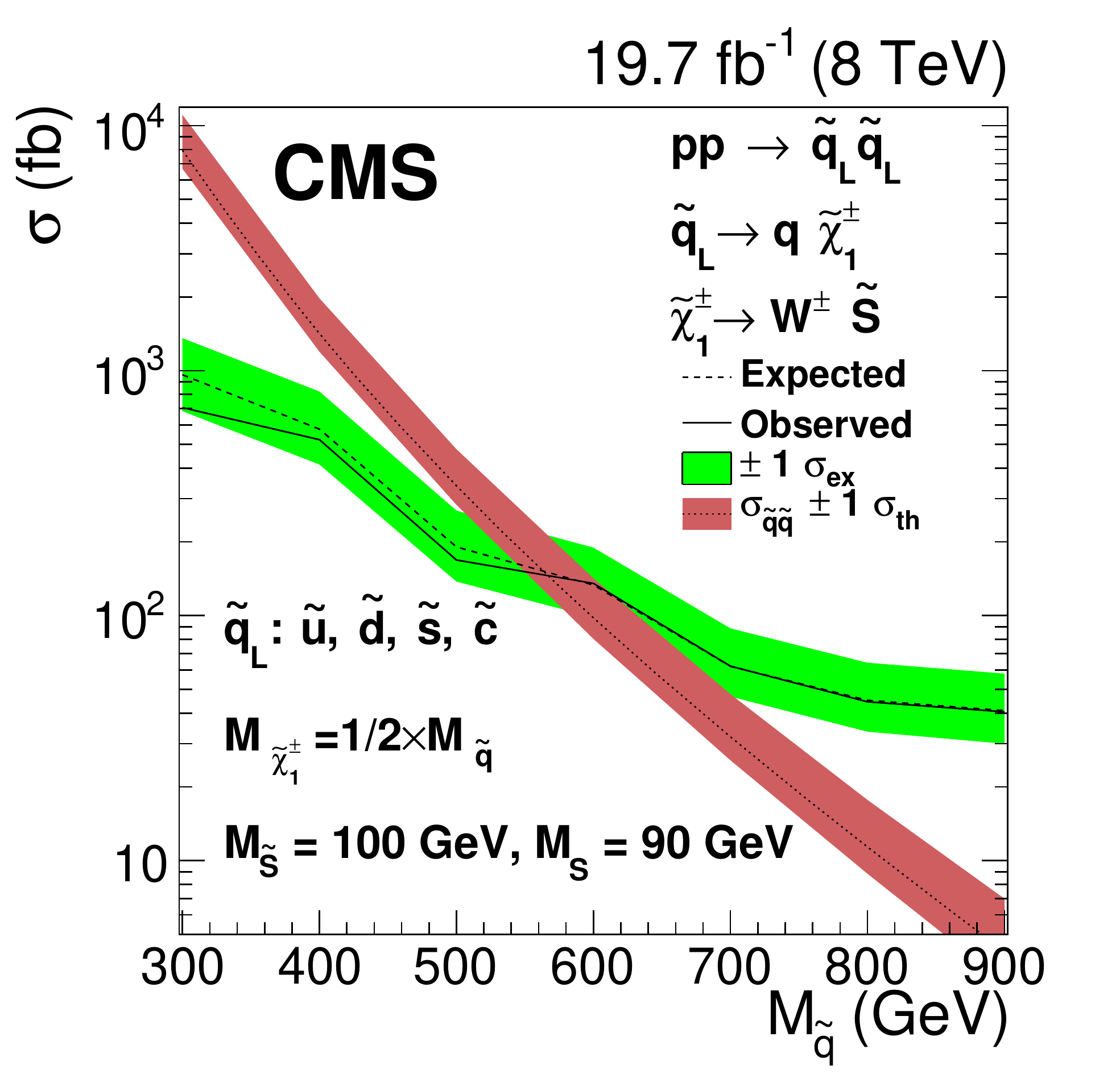}
  \caption{Observed and median expected cross section upper limits as a function of squark mass from
    the $\ell^\pm$ analysis. The band about the expected limit indicates the one standard deviation
    experimental uncertainty. The NLO+NLL cross section with its one standard deviation uncertainty
    is also shown.}
  \label{fig:limit_lepton}
\end{figure}

Figure~\ref{fig:totalLimit_doublePhoton} shows the cross section upper limits for the $\gamma$
analysis as a function of the squark and neutralino masses. The predicted NLO+NLL cross section is
used to place constraints on the masses of the squarks and neutralinos under the assumption of
stealth SUSY. We show the observed (median expected) mass exclusion with a band corresponding to the
variation of the theoretical (experimental) uncertainties by one standard deviation. For higher
neutralino masses, we exclude squark masses below 1050\GeV at a 95\% CL for the $\gamma$ analysis.
At low masses the neutralino becomes more boosted, and the resulting decay products are more tightly
collimated, spoiling the isolation of the photon. As a result the limit degrades for neutralino
masses below 300\GeV. Figure~\ref{fig:limit_lepton} shows the observed and median expected cross
section upper limits for the $\ell^\pm$ analysis as a function of squark mass for the model choices
described in Section~\ref{sec:mc}, as well as the predicted cross section from stealth SUSY. Based
on the intersection of the observed limit and the predicted cross section, we exclude squark masses
below 550\GeV at a 95\% CL.

\section{Summary}
\label{sec:sum}

We perform a search for new phenomena in events with four or more jets, low missing transverse
momentum, and either two photons ($\gamma$ analysis) or one electron and one muon of opposite charge
($\ell^\pm$ analysis), based on a data sample corresponding to an integrated luminosity of
19.7\fbinv of pp collisions at $\sqrt{s} = 8\TeV$. Using background estimation methods based on
control samples in data, we determine limits on the squark pair production cross section, and we use
those limits in conjunction with NLO+NLL cross section calculations to constrain the masses of
squarks and neutralinos in the framework of stealth SUSY. We do not observe a significant excess of
events above the standard model expectation in any search region. In the $\gamma$ analysis we
establish 95\% confidence level lower limits on squark masses between 700 and 1050\GeV, depending on
the neutralino mass. In the $\ell^\pm$ analysis we exclude squark masses below 550\GeV at the 95\%
confidence level. The mass limits for the $\gamma$ analysis supersede those from our previous
study~\cite{stealth_2012}. Our results for the $\ell^\pm$ analysis represent the first limits for
this channel.

\begin{acknowledgments}
We congratulate our colleagues in the CERN accelerator departments for the excellent performance of the LHC and thank the technical and administrative staffs at CERN and at other CMS institutes for their contributions to the success of the CMS effort. In addition, we gratefully acknowledge the computing centers and personnel of the Worldwide LHC Computing Grid for delivering so effectively the computing infrastructure essential to our analyses. Finally, we acknowledge the enduring support for the construction and operation of the LHC and the CMS detector provided by the following funding agencies: BMWFW and FWF (Austria); FNRS and FWO (Belgium); CNPq, CAPES, FAPERJ, and FAPESP (Brazil); MES (Bulgaria); CERN; CAS, MoST, and NSFC (China); COLCIENCIAS (Colombia); MSES and CSF (Croatia); RPF (Cyprus); MoER, ERC IUT and ERDF (Estonia); Academy of Finland, MEC, and HIP (Finland); CEA and CNRS/IN2P3 (France); BMBF, DFG, and HGF (Germany); GSRT (Greece); OTKA and NIH (Hungary); DAE and DST (India); IPM (Iran); SFI (Ireland); INFN (Italy); MSIP and NRF (Republic of Korea); LAS (Lithuania); MOE and UM (Malaysia); CINVESTAV, CONACYT, SEP, and UASLP-FAI (Mexico); MBIE (New Zealand); PAEC (Pakistan); MSHE and NSC (Poland); FCT (Portugal); JINR (Dubna); MON, RosAtom, RAS and RFBR (Russia); MESTD (Serbia); SEIDI and CPAN (Spain); Swiss Funding Agencies (Switzerland); MST (Taipei); ThEPCenter, IPST, STAR and NSTDA (Thailand); TUBITAK and TAEK (Turkey); NASU and SFFR (Ukraine); STFC (United Kingdom); DOE and NSF (USA).

Individuals have received support from the Marie-Curie programme and the European Research Council and EPLANET (European Union); the Leventis Foundation; the A. P. Sloan Foundation; the Alexander von Humboldt Foundation; the Belgian Federal Science Policy Office; the Fonds pour la Formation \`a la Recherche dans l'Industrie et dans l'Agriculture (FRIA-Belgium); the Agentschap voor Innovatie door Wetenschap en Technologie (IWT-Belgium); the Ministry of Education, Youth and Sports (MEYS) of the Czech Republic; the Council of Science and Industrial Research, India; the HOMING PLUS programme of Foundation for Polish Science, cofinanced from European Union, Regional Development Fund; the Compagnia di San Paolo (Torino); the Consorzio per la Fisica (Trieste); MIUR project 20108T4XTM (Italy); the Thalis and Aristeia programmes cofinanced by EU-ESF and the Greek NSRF; and the National Priorities Research Program by Qatar National Research Fund.
\end{acknowledgments}
\bibliography{auto_generated}

\cleardoublepage \appendix\section{The CMS Collaboration \label{app:collab}}\begin{sloppypar}\hyphenpenalty=5000\widowpenalty=500\clubpenalty=5000\textbf{Yerevan Physics Institute,  Yerevan,  Armenia}\\*[0pt]
V.~Khachatryan, A.M.~Sirunyan, A.~Tumasyan
\vskip\cmsinstskip
\textbf{Institut f\"{u}r Hochenergiephysik der OeAW,  Wien,  Austria}\\*[0pt]
W.~Adam, T.~Bergauer, M.~Dragicevic, J.~Er\"{o}, M.~Friedl, R.~Fr\"{u}hwirth\cmsAuthorMark{1}, V.M.~Ghete, C.~Hartl, N.~H\"{o}rmann, J.~Hrubec, M.~Jeitler\cmsAuthorMark{1}, W.~Kiesenhofer, V.~Kn\"{u}nz, M.~Krammer\cmsAuthorMark{1}, I.~Kr\"{a}tschmer, D.~Liko, I.~Mikulec, D.~Rabady\cmsAuthorMark{2}, B.~Rahbaran, H.~Rohringer, R.~Sch\"{o}fbeck, J.~Strauss, W.~Treberer-Treberspurg, W.~Waltenberger, C.-E.~Wulz\cmsAuthorMark{1}
% ...
\vskip\cmsinstskip
% ...
1:~~Also at Vienna University of Technology, Vienna, Austria\\
2:~~Also at CERN, European Organization for Nuclear Research, Geneva, Switzerland\\
% ...

\end{sloppypar}
\end{document}